\newcommand\sch{Schwarzschild}

\newcommand\text{Tr_{ext}}

\newcommand\gmn{g_{\mu \nu}}
\newcommand\beq{\begin{equation}}
\newcommand\eeq{\end{equation}}

\documentstyle[aps,prd]{revtex}
\begin{document}
\draft
\twocolumn[\hsize\textwidth\columnwidth\hsize\csname
@twocolumnfalse\endcsname

\rightline{SISSA-178/96/A}
\rightline{gr-qc/9612024}
\vskip1pc

\title{Black Hole Thermodynamics, Casimir Effect and Induced Gravity}
\author{F.~Belgiorno$^\ast$ and 
S.~Liberati$^\dagger$}
\address{$^\ast$Dipartimento di Fisica, Universit\`a di Milano, 20133 
Milano, Italy\\ 
E-mail address: belgiorno@vaxmi.mi.infn.it\\
$^\dagger$SISSA/ISAS, Via Beirut 3-4,
34100 Trieste, Italy\\
E-mail address: liberati@sissa.it}
\date{December 9, 1996}
\maketitle

\begin{abstract}
An analogy between the subtraction procedure in the Gibbons-Hawking 
Euclidean path integral approach to Horizon's Thermodynamics 
and the Casimir effect is shown. Then a conjecture about a 
possible Casimir nature of the Gibbons-Hawking subtraction is made in 
the framework of Sakharov's induced gravity. 
In this framework it  appears that the degrees of freedom 
involved in the Bekenstein-Hawking  entropy are naturally identified with 
zero--point modes of the matter fields. Some consequences of this 
view are sketched.
\end{abstract}

\vskip2pc
]

\section{Introduction}

Semiclassical Euclidean quantum gravity techniques play a key role 
in the investigation of thermodynamics of black holes. Nonetheless
their physical interpretation is still a matter of debate. 
In this paper we propose a new framework in which this 
interpretation is achieved in a natural way by focusing our attention on the
dynamics of quantum vacuum fluctuations in curved spacetime. We 
stress that this paper has a programmatic nature. 
We mean to approach a more quantitative 
level in a future work.\\

As a starting point for our reasoning, we shall summarize 
the path integral approach procedure following the steps 
first delineated by Gibbons and Hawking \cite{GH}. 
Given the classical Einstein-Hilbert action for gravity and the action of
classical matter fields, 
one formulates the Euclidean path integral by means of a Wick rotation. 
In a semiclassical (saddle point) expansion of the action around 
classical field configurations, one can find 
a non-trivial partition function 
even taking into account only the gravitational tree level action. 
Indeed, in the presence of static background manifolds with 
bifurcate Killing horizons, 
the requirement of non-singular behaviour for the 
solutions of the equations of motion implies the periodicity of imaginary time.
The period corresponds to the inverse Hawking-Unruh 
temperature because of the relation between the periodicity of the 
Euclidean Green's functions and the thermal character of the 
corresponding Green's functions in Lorentzian signature. 
Hence, in these cases, the effective action is truly 
a free energy function $\beta F$.

This method, schematically sketched above, contains a step not 
definitively understood, namely the ``reference" action subtraction 
for the gravitational tree level contribution. 
The action on shell consists of the usual Einstein-Hilbert action, of a 
surface term related to the extrinsic curvature and a Minkowskian 
subtraction term (the ``reference"  action). The latter 
is introduced by requiring that in flat spacetime the 
gravitational action is zero and it is necessary in order to obtain a 
finite value when evaluated on shell. In the following 
section we will further analyze this topic. 

\section{Gravitational action subtraction}

Hawking and Horowitz \cite{HH} have developed this subtraction scheme 
to the case of non-compact geometries. They considered the Lorentzian 
gravitational action for a metric $g$ and matter fields $\phi$:
\beq
I(g,\phi)=\int_{M} \left [{{R}\over{16\pi}}+L_{m}(g,\phi) \right ]+
{1\over{8\pi}}\oint_{\partial M} K.
\label{action}
\eeq
The surface term is needed to give rise to 
the correct equations under the constraint of fixed induced metric and matter 
fields on the boundary $\partial M$. 
The action 
is not well--defined for non-compact geometries: one has in the 
latter case to choose a rather arbitrary background $g_{0}, \phi_{0}$. Indeed 
Hawking and Horowitz chose a static background solution of the field equations. 
Their definition of the physical action is then:
\beq
I_{phys}(g,\phi)\equiv I(g,\phi)-I(g_{0},\phi_{0}).
\label{acfi}
\eeq
The physical action for the background is thus zero.
Further, it is finite for a class of fields $(g,\phi)$ 
asymptotically equal to $(g_{0},\phi_{0})$. For asymptotically flat 
metrics the
background is $(g_{0},\phi_{0})\equiv (\eta, 0)$; the action so obtained 
is then equal to that of Gibbons and Hawking:
\begin{eqnarray}
I_{phys}(g,\phi)&=&\int_{M} \left [ 
{{R}\over{16\pi}}+L_{m}(g,\phi) \right ]\cr
&&\cr
&+&{1\over{8\pi}} \oint_{\partial M} (K-K_{0}).
\label{giha}
\end{eqnarray}
The last term is just the Minkowskian subtraction: $K_{0}$ is the trace of 
the extrinsic curvature of the boundary of the background spacetime.
The above subtraction could be physically interpreted by requiring 
that it should represent that of a background contribution 
w.r.t. which a physical effect is measured.

There is a nontrivial point to be stressed about (\ref{acfi}): it 
is implicitly assumed that the boundary metric $h$ on $\partial M$ 
induced by $g_{0}$ and $g$ is the same. In general it is not 
possible to induce a 3--metric $h$ from a given 4--metric $g_{0}$; 
the same problem arises for the induction of a generic $h$ by 
flat space \cite{hawcen}. In the case where the 
asymptotic behaviour of the 4--metrics $g$ and $g_{0}$ is the same, one can 
assume that the 3--metrics, say $h$ and $h_{0}$, induced respectively by 
$g$ and $g_{0}$, become asymptotically equal \cite{hawcen}.
More generally the requirement to get the same boundary induced 
metric by $g_{0}$ and $g$ can be thought as a physical constraint 
on the choice of a reference background for a given spacetime.

So far we have argued that the subtraction procedure is a fundamental step 
in the path-integral formulation of semiclassical quantum gravity. In what 
follows, we will recall some well--known facts about the Casimir 
effect \cite{birrell,moste,PMG}, in order to suggest a formal 
similarity between Casimir subtraction and the above gravitational 
action subtraction.

\section{Casimir subtraction}

We start discussing the problem of two parallel infinite conducting 
plates; the energy density is obtained by means of the 
subtraction of the zero--point modes energy in absence of the planes from the 
zero--point mode energy in the presence of the two planes.  
One can in general formally define the Casimir energy as follows \cite{PMG}:
\beq
E_{casim}[\partial M]=E_{0}[\partial M]-E_{0}[0]
\label{caster}
\eeq
where $E_{0}$ is the zero--point energy and $\partial M$ is a 
boundary\footnote{In eq. (\ref{caster}) and in the following 
analogous equations concerning the Casimir effect a regularization of 
the right hand side terms is understood.}. 

Boundary conditions in the Casimir effect can be considered \cite{PMG} 
as idealizations of real conditions in which matter configurations or external 
forces act on a field. The most general formula for the vacuum energy is 
\beq
E_{casim}[\mbox{\boldmath $\lambda$}]=E_{0}[\mbox{\boldmath
$\lambda$}]-E_{0}[\mbox{\boldmath $\lambda_{0}$}]
\label{casgen}
\eeq
where $\mbox{\boldmath $\lambda$}$ is a set of suitable parameters
characterizing
the given configuration (e.g. boundaries, external fields,
nontrivial topology), and $\mbox{\boldmath $\lambda_{0}$}$ is
the same set for the configuration w.r.t. which the effect has to be 
measured.
In the case that $\mbox{\boldmath $\lambda$}$ represents an external field
$\bf A$, the vacuum
energy distortion induced by switching on the external field is given by:
\beq
E_{casim}[{\bf A}]=E_{0}[{\bf A}]-E_{0}[0].
\label{casfie}
\eeq
One can also take into account the finite temperature Casimir
effects\cite{plut,dowken}: in this case matter fields are not in their  
vacuum state, there are real quanta excited which are statistically
distributed according to Gibbs canonical ensemble.
The Casimir free energy is:
\beq
F_{casim}[\beta,\mbox{\boldmath $\lambda$}]
=F[\beta,\mbox{\boldmath $\lambda$}]-F[\beta,\mbox{\boldmath $\lambda_{0}$}].
\label{casfree} 
\eeq
The zero--point contribution\cite{plut} 
to the finite temperature effective action is 
simply proportional to $\beta$ (so it doesn't contribute to the 
thermodynamics).

The formal analogy of (\ref{acfi}) with e.g. (\ref{casfree}) 
consists just in the fact that in both cases there 
is a reference background to be subtracted in order to get a 
physical result. In particular, the subtraction (\ref{acfi}) 
is analogous to that in (\ref{casfie}); the obvious
substitutions being $\gmn$ in place of $A$ and $\eta_{\mu \nu}$ in place 
of $0$.\\
We stress that there are still substantial differences between 
(\ref{acfi}) and (\ref{casfie}) due to 
the fact that in (\ref{casfie}) the field $A$ is external whereas in 
(\ref{acfi}) the field $\gmn$ is the dynamical field 
itself; moreover, a deeper link of 
(\ref{acfi}) with the Casimir effect would require a quantum field whose 
zero point modes are distorted by spacetime curvature. 
Note that in the latter case one could naively invoke a Casimir 
effect w.r.t. the background spacetime $(M_{0},g_{0})$:
\begin{eqnarray}
F_{casim}[\beta,g]_{M}&=&F[\beta,g]_{M}-F[\beta,g_{0}]_{M_{0}}.
\label{casgra}
\end{eqnarray}
We stress that (\ref{casgra}) is purely formal and requires static 
manifolds $(M,g)$, $(M_{0},g_{0})$. For zero--temperature 
the idea underlying the Casimir effect, as seen above, 
is to compare vacuum energies in two physically distinct 
configurations. If the gravitational field plays the role of an external 
field, one can {\em a priori} compare backgrounds with different 
manifolds, topology and metric structure.
The non triviality one finds in defining meaningfully 
a gravitational Casimir effect can be easily understood for example 
in terms of the related problem of choice of the vacuum state for 
quantum fields \cite{birrell}. Moreover, in the presence of a 
physical boundary, the subtraction (\ref{casgra}) is ill--defined 
in general because the same embedding problems exist as for (\ref{acfi}). 
Despite these problems, we assume that it is possible to give a 
physical meaning to (\ref{casgra}). It is at least well know how 
to do this in the case of static spacetime with fixed metrics and 
topology like $R\times M^3$ where the spatial sections $M^3$ are 
Clifford-Klein space forms of flat, spherical or hyperbolic 3-spaces 
$(M^3=R^3/ \Gamma,S^3/ \Gamma, H^3/ \Gamma)$ 
\cite{DB,Ish,GB,GB2}\footnote{Here $\Gamma$ is the group of deck 
transformations for the given space\cite{Wolf}.}.
 
Then Sakharov's conjecture \cite{Sak} about the 
nature of the gravitational field represents a conceptual 
framework in which the analogy can be strongly substantiated.

\section{Induced gravity}

According to Sakharov's ideas, the Einstein--Hilbert gravitational action is 
induced by vacuum fluctuations of quantum matter fields and it 
represents a type of elastic resistance (of constant G) of the 
spacetime to being curved. The qualitative basis of this statement 
\cite{adler} is the fact that the Einstein--Hilbert action 
density is given by the Ricci scalar $R$ times 
a huge constant (order of the square of the Planck 
mass): curvature development requires a large action penalty 
\cite{adler} to be paid, that is there is an ``elastic" 
resistance to curvature deformations. The fact that a 
long--wavelength expansion of quantum matter fields in curved 
spacetime contains zero point divergent terms proportional to the 
curvature invariants\footnote{See also the following section.} 
according to Sakharov suggests that zero point fluctuations 
induce the gravitational action. ``Induction" means that no tree 
level action is considered: quantum matter fields generate it 
at a quantum level. Gravitational interaction in this picture 
becomes a residual interaction \cite{mtw} of a more fundamental 
one living at high energy scales (Planck mass); there 
are various ways to implement such a fundamental theory\cite{adln}.\\
The induced gravitational action should be 
given by the difference between the quantum effective zero--point 
action for the matter fields in the presence of the spacetime curvature and 
the effective action when the curvature is zero\cite{sakte} i.e.
\beq
S_{induced\ gravity}=\Gamma [R]-\Gamma [0].
\label{saka}
\eeq
The field $g_{\mu \nu}$ actually appears, in this low energy 
regime, as an external field and not as a dynamical one.
Then the induced gravity framework allows us to identify the 
Min\-kow\-skian subtraction as a Casimir subtraction. 
We note that there is a boundary term in (\ref{acfi}) that is 
necessary in order to implement a Casimir interpretation of the 
subtraction and that is missing in the original idea of Sakharov. 
But if the manifold has a boundary 
it is natural to take into account its 
effects on vacuum polarization\footnote{
In general a non trivial topology affects vacuum polarization 
\cite{moste}.}. As in a renormalization scheme for 
quantum field theory in curved spacetime it is necessary to 
introduce suitable boundary terms in the gravitational action in 
order to get rid of surface divergences\cite{KCD,birrell}, so 
there should be boundary terms in the induced gravitational action 
\cite{denardo}.  Anyway we don't know if taking into account 
boundary terms and suitable boundary conditions it is possible to produce a 
self--consistent theory of induced gravity\footnote{For a wider 
discussion on this point see also \cite{barv}.}. The choice of the boundary 
conditions should be constrained in such a way to get an induced 
gravity action with a boundary term as in Hawking 
approach. In this paper we shall limit 
ourselves to a discussion of the case of a scalar field and to the 
divergent part of the effective action.

In a curved manifold $M$ with smooth boundary $\partial M$, 
the zero--point vertex functional for a scalar field 
depends on the curvature and is divergent:
\beq
\Gamma [\phi=0,\gmn]=\Gamma [\gmn].
\label{zercur}
\eeq
The zero--point effective action (\ref{zercur}) is comprised of  divergent 
terms that, if $D$ is the dimension of $M$, correspond to the first 
$l \leq D/2$ ($l=0,1/2,...$) coefficients, $c_{l}$, in the heat kernel 
expansion \cite{KCD,dowken}. In our case $D=4$ and $l \leq 2$. 
For a smooth boundary the coefficients $c_{l}$ 
can be expressed as a volume part plus a boundary part:
\beq
c_{l}=a_{l}+b_{l}.
\eeq 
The $b_{l}$ depend on the boundary geometry and on the boundary 
conditions. The $a_{l}$ coefficients vanish for $l$ half integral and for 
integral values are equal to the Minakshisundaram coefficients for the 
manifold $M$ without boundaries. 

If there is a classical (not induced but fundamental) gravitational 
action, the divergent terms in (\ref{zercur}) can be renormalized\cite{KCD} 
by means of suitable gravitational counterterms: that is by reabsorbing 
the divergences into the bare gravitational constants 
appearing in the action for the gravitational field: 
\beq
S_{ren}[\gmn]=S_{ext}[\gmn]+\Gamma_{div} [\gmn].
\label{reno}
\eeq
In an induced gravity framework, there is no classical (tree 
level) term like $S_{ext}[\gmn]$ to be renormalized and so there should 
exist a dynamical cut--off rendering finite also the divergent 
terms. These terms give rise to the gravitational (effective) 
action, so we can call $\Gamma_{div} [\gmn]$ ``the gravitational part'' 
of the effective action.

In the case of finite temperature field theory on a static 
manifold with boundary, the standard periodicity condition in the 
imaginary time $\tau$ with period $\beta$ can be implemented by 
means of the following image sum over a non--periodic heat kernel:
\beq
K_{\beta}(x,y;s)=\sum_{n=-\infty}^{+\infty} K_{\infty}(x,y-n k 
\beta; s)
\label{kabe}
\eeq
where $s$ is the usual ``fifth coordinate" and $k$ is a four 
vector in the same direction as the periodic coordinate. 
The calculation of the partition function of the matter fields 
is meant to be carried out in the so called ``on shell" 
approach\cite{fste}, i.e. without introducing any conical defect in 
the manifold\footnote{In this way no further divergence on the 
horizon is expected.}. 
The $n=0$ term in (\ref{kabe}) is ordinarily a zero 
temperature--term and it is the only divergent one (cf. 
\cite{dowken}). It corresponds, in the induced gravity framework, to the 
gravitational contribution. 

We are mainly interested in the case of the \sch\ black hole: 
(\ref{acfi}) could be then interpreted as a Casimir free 
energy contribution relative to the matter field zero--point modes. 
We consider the one--loop divergent 
contribution for a massive scalar field enclosed in a sphere with radius 
$r=R_{box}$.

We choose the boundary 
condition by looking at the structure of the boundary terms. 
For consistency, one would get in particular the 
Einstein--Hilbert action term (\ref{acfi}) at the same order in 
the heat kernel expansion, so we point our attention to the 
boundary term $b_{1}$. 
It is possible to get the right form for 
the integrated coefficient 
\beq
\int_{\partial M} K
\eeq
both for Dirichelet and Neumann boundary conditions. 
The first one is selected on the physical grounds that for a 
sufficiently large box (infinite in the limit) the field should 
be zero on the boundary. 

Of course in order to get ordinary 
gravitational dynamics, i.e. General Relativity, it is necessary that the 
couplings and the mass of the fundamental theory fulfill 
suitable renormalization constraints. 

In our conjecture, the gravitational part of the 
free energy (\ref{acfi}) becomes a Casimir free energy 
contribution arising by zero--point modes. 
What should it mean from a physical point of view? 
The most naive answer to this question is that 
black hole (equilibrium) thermodynamics becomes a thermal 
physics of quantum fluctuations that are initially in a spherically 
symmetric spacetime 
and then are thermally distorted by the formation of a black hole. 
In this view it seems that there is implicitly an idea of a 
``physical process" underlying the integral version 
of the black hole thermodynamics laws, i.e. one has to take into 
account that the vacuum \sch\ solution has been generated by a 
gravitational collapse drastically deforming the vacuum of the quantum 
fields vacuum\footnote{We underline that the idea of a link between the Casimir 
effect, induced gravity and black hole evaporation was formulated by 
U.H. Gerlach\cite{Gerl} in a model for an incipient black 
hole where a sort of dynamical Casimir effect prevents the 
formation of the event horizon.}. 
Also in the general case, we conjecture the same 
Casimir interpretation for 
the subtraction relative to the the zero--point terms 
(gravitational Lagrangian) in the effective action. 
Outside of the induced gravity framework, 
if one naively considers a generalized Casimir effect 
for quantum fields, e.g. on the \sch\ background w.r.t 
flat space, one gets a zero--point contribution to be renormalized in the 
gravitational action and, for consistency with (\ref{casgra}), 
the gravitational action has to follow the Casimir subtraction 
scheme. What one would miss in this case is a microscopic interpretation 
of the tree level gravitational contribution.

\section{Thermal behaviour of zero point modes}
The induced gravity framework implies that black hole entropy can by 
explained 
in terms of Boltzmann's counting of microstates by identifying the statistical 
mechanical degrees of freedom with the zero--point fluctuations of quantum 
matter fields\footnote{This idea is pursued with different 
conceptual tools in the papers of Jacobson\cite{Jacpre}, of 
Frolov, Fursaev and Zelnikov\cite{Froze} and also of 
Gerlach\cite{Gerl}. Zero--point modes 
could explain more generally the entropy appearing in 
Horizon Thermodynamics.}. This particular view seems nonsense in the 
framework of standard (i.e. without the Hawking--Unruh effect) 
statistical mechanics, because in this case zero--point modes cannot 
contribute to the entropy, their contribution to the effective action being 
proportional to $\beta$. But there is no real contradiction: 
in the case of horizon's thermodynamics, the subtracted 
gravitational action is not simply proportional to $\beta$. In black holes 
spacetime it is proportional to $\beta^{2}$. In this case there is a bifurcate 
Killing horizon that is related to the 
presence of thermal zero--point mode contribution. 
This means that vacuum fluctuations do contribute to the 
entropy. Note that this conclusion is independent of
the induced gravity framework: matter fields 
give a thermal zero--point contribution (that has to be renormalized in the
gravitational action outside of Sakharov's viewpoint).

The topological structure of spacetime seems to be deeply 
linked to the thermal behaviour of quantum fluctuations. 
They are expected to have stochastic properties 
described by the structure of the n-th order correlation 
functions; a thermal spectrum is a subcase requiring 
both a Gaussian distribution (n-order correlations zero for n$\geq 3$)
and a Planckian frequency spectrum (the Fourier transform of the 
2-point correlation function)\cite{SCD}. A bifurcate Killing horizon in a 
static manifold seems to ensure a Gaussian behaviour that in other more general 
cases (e.g. non static manifolds without bifurcate Killing 
horizons) is substituted by a ``standard" stochasticity.  

In order to see that it is possible to find a thermalization the zero--point 
modes, we recall some facts concerning the Unruh 
effect in Rindler spacetime. 
An uniformly accelerated observer in a Rindler wedge 
of Minkowski spacetime perceives Minkowski vacuum as a thermal 
state satisfying a KMS condition, that is a detailed balance condition. 
The ``particle" spectrum seen by the accelerated observer depends only on 
the acceleration and not on the velocity of the observer, and this 
fact means that actually zero--point fluctuations\cite{SCD} of 
Minkowski vacuum appear as particles thermally 
distributed in the accelerated detector\footnote{See also 
\cite{sarm} and references therein.}. 
Moreover, at a formal level, the thermofield dynamics approach\cite{TU}
can explain why a pure state (Minkowski vacuum) appears 
as a mixed one (thermal) to a Rindler observer\cite{walbo}: 
there is a horizon hiding part of the information relative to the state. 
A similar thermofield scheme is also available for the \sch\ black 
hole\cite{walbo,Werner}, in which the information relative to a vacuum 
state defined on the Kruskal extension (the Hartle--Hawking 
state) appears to be thermal to a static observer in the 
\sch\ external region. 
Actually, it has to be stressed that the thermofield framework can be used, 
only if one can consider the maximal extension of a spacetime and even in 
these cases one can find, as in Kerr black holes (cf. \cite{walbo}), that 
it is impossible to mimic the Unruh-Rindler scheme. 
So from the Unruh effect we have some insights about zero--point mode 
thermalization, but no general and definitive 
explanation available also in the cases of black holes arising from a 
realistic collapse, rotating. 

A more general explanation of vacuum thermalization can be found in the 
framework of non-equilibrium horizon thermodynamics introduced 
by Sciama\cite{DWS}. In Sciama's approach the irreversible 
process of black hole evaporation can be interpreted as a progressive 
dissipation of the geometry against vacuum fluctuations. 
In his view, black hole and radiation are linked by a
fluctuation-dissipation relation for their zero-point modes. 
In the static case a local equilibrium condition is achieved 
in which the only distribution of the quantum states 
that is independent of time and stable w.r.t weak interactions between 
the modes of the field does satisfy the KMS condition 
(i.e. it exists a thermal distribution). 
This is possible apparently only in presence of proper degrees 
of freedom for the black hole. As an atom can come into equilibrium with 
vacuum fluctuations of the electromagnetic field 
(its zero-point fluctuations being ``enslaved'' by those of the field) 
so the black hole would come into equilibrium only if endowed 
with proper degrees of freedom. Hence, in order to pursue the above 
analogy, one has to deal with the highly non trivial 
problem of identifying a physically acceptable notion of 
degrees of freedom for a black hole, a problem that is commonly believed to be 
associated to a full quantum gravity theory rather than to a semiclassical 
approach. 

The interpretation we are proposing of gravitational action 
and horizon thermodynamics appears as a strong link to Sciama's framework. 
Indeed it describes the geometry-matter system at the 
equilibrium as a sort of static Casimir effect and the non-equilibrium 
regimes are 
naturally interpreted as dynamical Casimir-like effects in the same way as 
the energy spent by the geometry in thermally exciting vacuum fluctuations 
can be interpreted in Sciama's view as an infalling of negative energy into the 
black hole. Moreover, in our framework of induced gravity, 
black hole entropy is associated with a Casimir distortion of 
vacuum fluctuations. 
Finally, regarding the problem of a fluctuation--dissipation theorem 
for black holes, our proposal leads us to conjecture that one could bypass 
the request of internal degrees of freedom for the black hole (at least 
at this mesoscopic level) as follows. Geometry deformations w.r.t. the 
static case (like quasinormal modes of black hole\cite{chayork}) should be 
driven to a stationary state (static black hole). 
This means that the way geometry affects the quantum matter field spectrum 
should be such that the static black hole geometry, although quantum 
unstable (through Hawking radiation), is more stable than a deformed black 
hole geometry. 

\section{Conclusions}

We conclude with some comments on the implications of our framework.
Firstly it can be cast in the recently sustained interpretation of 
General Relativity as an effective theory\cite{Jac,Hu}. In Sakharov's picture
the Einstein--Hilbert action, being induced by quantum matter
field fluctuations, should not be considered
as the action of a fundamental theory. It can be interpreted
as the (effective) long--wavelength 
action of a mesoscopic theory, as elasticity
is a mesoscopic theory which is related to the fundamental theory
of quantum electrodynamics. So quantizing gravity could be equivalent
to quantizing phonons \cite{Hu}.

Our attempt consists in finding a link with Casimir 
physics suggested by the subtraction procedure in 
black hole thermodynamics. The subtraction in our view 
takes into account a physical process of adiabatic vacuum energy 
distortion and in a real collapse we expect 
nonadiabatic contributions.  
To be more specific about this point, we shall 
shortly summarize the general framework we choose.

Quantum matter fields coupled with geometry 
induce geometrodynamics action terms that correspond to zero 
point fluctuations and are both local (bulk part) and global 
(boundary part). One can also look for a more fundamental 
pregeometric theory living at the planckian scale 
(e.g. noncommutative geometry, string theory). 
In any case, the low energy theory gets from high energy theory only a 
dynamical scale and the value of the couplings (via the renormalization 
group). In Sakharov's view, one can conclude that the 
degrees of freedom involved in the gravitational action are vacuum 
fluctuations that can explain black hole entropy\cite{Jacpre}. 
In particular the vacuum acts as a viscous (dissipative) medium whose 
distortion, due to a non trivial topology and/or geometry configuration, 
gives rise to black hole entropy. 
Note that the Casimir interpretation of (\ref{acfi}) involves only 
external degrees of freedom i.e. external vacuum fluctuations 
without involving any notion of internal states. In particular 
static black hole entropy is a static ``Casimir" entropy. 
In a dynamical configuration such as gravitational collapse or black hole 
evaporation a dynamical Casimir effect would be involved.

About the prescription one has to follow in order to actually compute 
such an entropy, our framework implies a conceptually easy prescription.
The gravitational free energy is now the free energy for the matter 
field zero point modes. Once one calculates the latter one can find the 
correspondent entropy by applying the usual formula $F=\beta E-S$.
In \sch\ case the internal energy is the black hole mass.

Finally, we think that an induced gravity framework could explain 
why General Relativity, as a classical theory, knows about a quantum 
phenomenon like the Hawking effect. Indeed, the so called four laws 
of black hole mechanics were found at a classical level before of 
the discovery the Hawking effect (quantum level). Quantum 
radiation from black holes has given a semantic meaning to a 
surprising syntactic analogy between black hole mechanics and 
standard thermodynamics. But the meaning of thermodynamical behaviour of the 
classical level gravity is still mysterious. 
We think that imposing Einstein's gravity to be an effect of order 
$\hbar$ could represent a good bridge to such an understanding. 

Of course, there are many open questions to be solved. 
Quantitative and further conceptual developments of our approach are 
deferred to a future publication.

\section*{Acknowledgements}

The authors are particularly indebted with
D.W.Sciama for several illuminating discussions.
They wish to thank B. Bassett, S.Sonego, G.Immirzi, L.Pilo, G.Pollifrone,
K. Yoshida and M.Martellini for useful comments and remarks.

\end{document}